\documentclass[a4paper,10pt]{article}

\usepackage[english]{babel}
\usepackage[utf8]{inputenc}
\usepackage{amsmath}
\usepackage{graphicx}
\usepackage[colorinlistoftodos]{todonotes}
\usepackage{verbdef}
\usepackage{titling}

\usepackage[bookmarks, bookmarksopen, bookmarksnumbered, hidelinks, pdfborder={0 0 0}]{hyperref}
\usepackage[all]{hypcap}
\usepackage{xcolor}
\hypersetup{
    colorlinks,
    linkcolor={red!50!black},
    citecolor={blue!50!black},
    urlcolor={blue!80!black}
}

\verbdef{\jsonexample}|{ "name": "Daniel S. Katz" } |
\verbdef{\name}|"name" |
\verbdef{\keywords}|"keywords" |
\title{Implementing Transitive Credit with JSON-LD \vspace{-0.2cm}}

\author{
Daniel S. Katz\\
National Science Foundation;\\
Computation Institute, University of Chicago \&\\
Argonne National Laboratory\\
\url{d.katz@ieee.org}
\and
Arfon M. Smith\\
GitHub Inc.\\
\url{arfon@github.com}
}

\date{}

\begin{document}

\maketitle

\vspace{-1.2cm}
\begin{abstract}

\noindent
Science and engineering research increasingly relies on activities that facilitate research but are not currently rewarded or recognized, such as: data sharing; developing common data resources, software and methodologies; and annotating data and publications.   
To promote and advance these activities, we must develop mechanisms for assigning credit, facilitate the appropriate attribution of research outcomes, devise incentives for activities that facilitate research, and allocate funds to maximize return on investment. In this article, we focus on addressing the issue of assigning credit for both direct and indirect contributions, specifically by using JSON-LD to implement a prototype transitive credit system.
\end{abstract}

\section{Introduction}

Science and engineering research increasingly relies on activities that facilitate research but are not currently rewarded or recognized. This includes the sharing of data; development of common data resources, software and methodologies; and annotation of data and publications. This situation has been documented in a number of recent reports~\cite{NSF_software_vision, WSSSPE1} that focus on changing needs and mechanisms for attribution and citation of digital products, from the use of alternative metrics~\cite{altmetrics} that track reports of research impact apart from research publications, to work on data~\cite{nrc_data}. About half of the articles in many recent issues of Science describe research that depended on software, and a larger fraction analyze data. Indeed, the US National Science Foundation recently updated its guide to proposers to instruct them to provide a list of their ``products''---objects that are ``citable and accessible including but not limited to publications, data sets, software, patents, and copyrights''---rather than publications~\cite{nsf_gpg}.  

To promote and advance pursuit of activities that facilitate research, we must develop mechanisms for assigning credit, facilitate the appropriate attribution of research outcomes, devise incentives for activities that facilitate research, and allocate funds to maximize return on investment. In this article, we explore how the idea of transitive credit~\cite{Katz2_WSSSPE, Katz2_WSSSPE_JORS}, which would credit both direct and indirect contributions, can be implemented.

\section{Transitive Credit}

Transitive credit involves three main elements.  The first is complete credit. Any product should list all authors (as currently listed as authors of a paper), all contributors (as currently listed in the acknowledgements of a paper) and all component products that have been used, including both publications and other products such as software and data (as currently either cited, acknowledged, or not included in a paper). We coin the term ``contriponents'' for the combination of these contributors and components (though we eagerly welcome suggestions for a better term).

Second, all the contriponents need to have weights assigned.
Determining how to weight credit of the authors may be difficult, but it should be possible. 
Methods for doing this weighting, whether using a taxonomy or a more traditional list of authors, and analysis of these methods and their impact would likely happen if this overall idea moves forward.
Just as publications today are submitted by one person who is responsible for making sure all authors are listed (and perhaps assigned roles in a taxonomy) and the publication is complete, the submitter would also be responsible for registering this fractional credit, no matter how the values are determined.

The third element of transitive credit is its transitive nature, as is shown, for example, by how the credit map for a product A, which is used by a product B, feeds into the credit map for product B. Suppose product A is a software package equally written by two authors and its credit map is that 50\% of the credit for this should go the lead developer, 20\% to the second developer, and 10\% to the third developer. In addition, 5\% should go to each of the four libraries that are needed to run the code. When this product is created and registered, this credit map is registered along with it.  Product B is a paper that obtains new science results, and it depended on Product A. The person who registers the publication also registers its credit map, in this case 75\% to her/himself, and 25\% to the software code previously mentioned. Credit is now transitive, in that the lead software developer of the code can be given credit for 12.5\% of the paper. If another paper is later written that extends the product B paper and gives 10\% credit to that paper, the lead software package developer will also have 1.25\% credit for the new paper.

The value of transitive credit is in measuring the indirect contributions to a product, which today are not quantitatively captured. Because they aren't captured, they aren't rewarded, and there is a disincentive to perform them, due to the cost (in time or something else). If they were captured, this disincentive would be replaced by an incentive, which for software and data would mean to publish and share them in a reusable form.

Transitive credit could be implemented by adding either creditmaps to the metadata stored with DOIs~\cite{DOI-data-model}, or alternatively by building a separate system to store creditmaps and then adding pointers to entries in this system to the metadata stored in DOIs. Here, we focus on the creditmap itself, and how it can be described in a structured, machine (and human) readable form.

\section{JSON-LD}

JSON-LD (JavaScript Object Notation for Linked Data, \url{http://json-ld.org}) is a subset of the key-value based JSON document format that provides a way of describing machine-readable information with semantic context. Popular as an alternative to XML in the web development community, JSON is also used as a base data format for search engines such as Elastic Search~\cite{elasticsearch} and NoSQL data stores such as MongoDB~\cite{mongodb}.

In the final stages of standardization at W3C~\cite{w3c}, JSON-LD is designed to lower the barrier for data publishers who wish to provide `Linked Data' so that concepts and entities can be identified with certainty. As an example of machine readable data with semantics take this JSON snippet describing a person: \jsonexample . Without additional \textit{context} there is ambiguity as to what the \name term describes (such as the name of a person, place or thing). A better alternative is:

{\footnotesize
\begin{verbatim}
{
  "@context": "http://schema.org",
  "@type": "Person",
  "name": "Daniel S. Katz",
  "@id": "http://orcid.org/0000-0001-5934-7525"
 }
\end{verbatim}
}

\section{Using JSON-LD for Transitive Credit}

Smith recently proposed using JSON-LD for research tools~\cite{arfon-json-ld-blog}. In this paper, we extend this idea to suggest that it could be used for transitive credit of any scholarly product. Because of namespaced nature of the JSON-LD structure, it is trivial to include all contriponents such as datasets, software, and articles with their appropriate semantic context definition while maintainting a both human and machine-readable structure. Note that we use vocabularies from \url{http://schema.org} for the different entities; other examples include DOAP (\url{https://github.com/edumbill/doap}) and SPDX (\url{http://spdx.org}).

A subset (due to space limits) of a possible creditmap for this article follows, as an illustrative example of the power of this concept. 

{\footnotesize
\begin{verbatim}
{
  "@context": "http://schema.org",
  "@type": "ScholarlyArticle",
  "headline": "Implementing Transitive Credit with JSON-LD",
  "dateCreated": "2014-07-10",
  "keywords": "transitive credit, credit for code, json-ld, linked data",
  "author": [
    {
    "@type": "Person",
    "name": "Daniel S. Katz",
    "@id": "http://orcid.org/0000-0001-5934-7525",
    "email": "d.katz@ieee.org"
    "creditWeight": "0.25"
    },
    {
    "@type": "Person",
    "name": "Arfon Smith",
    "@id": "http://orcid.org/0000-0002-7217-4494",
    "email": "arfon@github.com",
    "creditWeight": "0.25"
    }
  ],
  "citation": {
    "articles": [
      {
        "@type": "ScholarlyArticle",
        "headline": "Transitive credit as a means to address social and technological 
            concerns stemming from citation and attribution of digital products",
        "doi": "10.5334/jors.be",
        "creditWeight": "0.3"
      }
    ],
    "software": [
      {
        "@type": "Code",
        "name": "Fidgit",
        "codeRepository": "https://github.com/arfon/fidgit",
        "license": "http://opensource.org/licenses/MIT",
        "creditWeight": "0.04"
      }
    ],
    "acknowledgment": [
      {
        "@type": "Person",
        "name": "James Howison",
        "@id": "http://orcid.org/0000-0002-5702-149X",
        "email": "james@howison.name",
        "creditWeight": "0.01"
      }
    ],
    "other": [
      {
        "@type": "BlogPosting",
        "headline": "JSON-LD for software discovery, reuse and credit",
        "url": "http://www.arfon.org/json-ld-for-software-discovery-reuse-and-credit",
        "license": "http://creativecommons.org/licenses/by/4.0/",
        "creditWeight": "0.15"
      }
    ]
  }
}
\end{verbatim}
}

\section{Conclusions}

In this paper we have outlined a mechanism for ascribing complete credit for all contriponents that have led to a scholarly product.

This is one of the needed elements to a transitive credit system.  This credit information could be stored as part of the metadata within the DOI system, perhaps by updating the kernel metadata and adding a creditmap entry to the DOI data dictionary~\cite{DOI-data-model}, though note that this is a standardization activity which requires community effort, similar to that being undertaken today by CASRAI (\url{http://casrai.org}) and others. Then, creditmap information would need to be added by registration authorities (RAs) that hold these scholarly products (i.e., publishers of all sorts, including journals, data archivers).  And it needs to be supported by indexing systems (e.g., Thomson Reuters). Similar complementary changes have also been discussed recently aimed at characterizing contributions rather than assigning  weights~\cite{metatags,contributorship-taxonomy,teams}.

The manner in which we use unique IDs is somewhat rough; we use ORCIDs for people and DOIs for papers, but for other elements, such as software that may not have a DOI, the solution is not clear.
Additionally, the manner in which weights are assigned is also rough; it might be easier to assign weights within categories, then assign weights to categories, and let the system determine the detailed weights.

While this paper addresses the technical aspects of how to make transitive credit possible, many social questions remain unresolved. For example, disciplinary communities will need to decide what the contriponents are that are relevant to their discipline, how to weigh various categories of contriponents (e.g., are developers more, equally, or less important than libraries?), and how the authors of products should assign weights to specific product contriponents (e.g., reference 1 is more/less important to a manuscript than reference 2.) 

One potential extension to this creditmap we have proposed would be to replace the~\keywords entry to include a description of the subject area from a defined taxonomy. When appropriate, such as in the case of software, this could also include a description of the function of the tool. Using indexing tools such as Elastic Search~\cite{elasticsearch}, it would then be possible to build indexes of these fields and then make powerful faceted searches such as `find astrophysics software, written in Python designed to manipulate spectroscopic data'.

Finally, we note that our proposal in this paper is very compatible with the current altmetrics activities~\cite{altmetrics}, and would add to the alternative metrics that could be collected and analyzed.

\section*{Acknowledgments}

The authors thank James Howison, Gabrielle Allen, and David Proctor for some discussions about this work.

Some work by Katz was supported by the National Science Foundation (NSF) while working at the Foundation; any opinion, finding, and conclusions or recommendations expressed in this material are those of the author and do not necessarily reflect the views of the NSF.

\bibliographystyle{plain}
\bibliography{wssspe_paper}

\end{document}